\documentclass[a4paper,12pt]{article}
\usepackage{amsmath,amsfonts,amssymb,cite}

\newcommand{\be}{\begin{equation}}
\newcommand{\ee}{\end{equation}}
\newcommand{\ba}{\begin{array}}
\newcommand{\ea}{\end{array}}
\newcommand{\bea}{\begin{eqnarray}}
\newcommand{\eea}{\end{eqnarray}}

\begin{document}

\begin{center}
{\Large\bf A holographic relation between the deconfinement temperature and gluon condensate}
\end{center}

\begin{center}
{\large S. S. Afonin\footnote{E-mail: \texttt{s.afonin@spbu.ru}}}
\end{center}

\begin{center}
{\it Saint Petersburg State University, 7/9 Universitetskaya nab.,
St.Petersburg, 199034, Russia}
\end{center}

\begin{abstract}
We derive a holographic prediction of the deconfinement
temperature $T_c$ at vanishing chemical potential within a simplest AdS/QCD
model with dynamical dilaton. Our analysis leads to a linear
relation between $T_c^4$ and the gluon condensate. After normalizing
this relation to the lattice data for $SU(3)$ pure gauge theory,
the standard phenomenological value of gluon condensate
from QCD sum rules leads to the prediction $T_c=156$~MeV which is
in a perfect agreement with the modern lattice results and
freeze-out temperature measured by the ALICE Collaboration.
\end{abstract}

\section{Introduction}

Quantum Chromodynamics (QCD) predicts that at high temperatures and/or hadron densities strongly
interacting matter exhibits a transition which separates the hadronic, confined phase and the quark-gluon plasma (QGP)
phase. The quantitative mapping of QCD phase diagram stays among the major challenges of the physics of
strong interaction. The ongoing heavy ion collision experiments at RHIC at Brookhaven National Laboratory, ALICE and SPS at
the Large Hadron Collider (LHC) are trying to quantify the properties of the deconfined phase
of strongly interacting matter --- the phase in which our early universe existed.
A special interesting region on the phase portrait is situated
near the onset of deconfinement transition at vanishing or small baryon (or quark) chemical
potential. The matter is that this region can be studied directly
from QCD using lattice simulations and the corresponding critical
temperature $T_c$ for deconfinement transition can  be calculated.
The two leading lattice collaborations in this field reported the values
$T_c=156\pm9$~MeV~\cite{Borsanyi:2010bp} and $T_c=154\pm9$~MeV~\cite{Bazavov:2014pvz}.
At physical quark masses $T_c$ represents in fact
a pseudo-critical temperature of crossover region between the hadron
phase and QGP. From the experimental side, $T_c$ is believed to be very close to
the temperature of chemical freeze-out $T_{cf}$ which is extracted from
a thermal analysis of the data on relativistic
nuclear collisions. The recent precise measurement by
ALICE Collaboration at LHC resulted in the value
$T_{cf}=156.5\pm1.5$~MeV~\cite{Andronic:2017pug}. Thus $T_c$ from lattice
simulations and $T_{cf}$ remarkably agree within errors.

From the theoretical viewpoint, however, the lattice calculations
represent a kind of "black box" whose results cannot be checked
analytically. On the other hand, the strict relation between $T_c$ and $T_{cf}$
is an open problem. It is therefore highly desirable to have
independent estimates for $T_c$ from models which were successful
in description of QCD phenomenology. A strong point of these complementary
approaches is that they can be able to relate $T_c$ to some important
dimensional quantity from low-energy QCD phenomenology. Looking from the opposite
side, a correct prediction of the value of deconfinement temperature
represents an important test for any model of this kind.

The deconfinement transition in QCD is a non-perturbative phenomenon occurring
at strong coupling. A promising modern theoretical tool for dealing with
strongly coupled gauge theories represents the idea of gauge/gravity duality~\cite{mald,witten}.
The application of this idea to low-energy phenomenology of QCD, started in Refs.~\cite{hw,sw},
turned out to be unexpectedly useful. The constructed bottom-up holographic models link
several successful approaches --- QCD sum rules, vector meson dominance, and
chiral perturbation theory --- into one framework through the gauge/gravity duality.

The deconfinement transition is described in holographic models as
a first order Hawking-Page type phase transition~\cite{hawking} between two different gravitational backgrounds ---
the thermal Anti-de Sitter (AdS) space and the Schwarzschild-AdS black hole~\cite{Witten:1998zw}. This idea was used
by Herzog in Ref.~\cite{Herzog} to calculate $T_c$ within the framework of two simplest holographic
QCD models --- the Hard Wall (HW)~\cite{hw} and Soft Wall (SW)~\cite{sw} model. The phenomenological value
of $T_c$ was reproduced with about 20\% accuracy
($T_c=122$~MeV for HW and $T_c=191$~MeV for SW model~\cite{Herzog}). The given analysis triggered a large
activity in the field. Perhaps the main advantage of Herzog's calculation is its
remarkable simplicity. On the other hand, this calculation used rather crude approximations.
First, the parameters of models were normalized to the $\rho$-meson mass although the
original actions did not contain something related to real hadrons. A more consistent
interpretation would be to consider this analysis as a calculation of deconfinement temperature
in pure gluodynamics, let us denote it as $T_c^{\text{gl}}$, since
the thermodynamics was assumed to be governed by the gravitational part of the action. Indeed, one can show that if
parameters of HW and generalized SW models are normalized to the mass of lightest glueball from
lattice simulations then the lattice value
$T_c^{\text{gl}}\simeq260\pm10$~MeV~\cite{Boyd:1996bx,Iwasaki:1996ca,Lucini:2012wq}
is reproduced in both approaches~\cite{estim_deconf}.
In addition, the deconfinement phase transition is of first order in the $SU(N)$
gluodynamics at $N\geq3$ and becoming increasingly abrupt with growing $N$~\cite{Lucini:2013qja}.
This means improving consistency: In the limit $N\rightarrow\infty$,
the phase transition becomes of the same type as
the Hawking--Page first order transition.

The second weak point is that both HW and SW holographic models are not
solutions of Einstein equations and this drawback is inherited in many other holographic
calculations of $T_c$ followed by Ref.~\cite{Herzog}. The reason lies in a need to introduce
a finite scale associated with the fifth direction in order to describe the Hawking-Page transition
for infinite boundary volume. The corresponding scale (associated with
the confinement scale) is inserted into the HW model via a hard cut-off of holographic
coordinate and into SW model through a certain static dilaton background.

In this Letter, we propose a somewhat new scheme for holographic calculation
of $T_c$. Our analysis retains a conceptual simplicity of Herzog's calculation
for the HW holographic model
but is free from aforementioned crude approximations and closer to phenomenological
gluodynamics. The main proposal consists in replacing the empty thermal AdS$_5$ space with a hard
cut-off imposed on the holographic coordinate by the simplest gravity-dilaton system in AdS$_5$
whose analytical solution
is known and where the cut-off emerges dynamically. The dilaton in holographic description of QCD
is usually associated with a field dual to the gluon condensate $\langle G_{\mu\nu}^2\rangle$ ---
an important phenomenological quantity parametrizing the mass gap in gluodynamics that appears
due to a dynamical violation of scale invariance in massless QCD and measures the QCD vacuum energy density.
We will get a simple relation between $T_c$ and
$\langle G_{\mu\nu}^2\rangle$ and demonstrate its phenomenological viability.

Physically the gluon condensate in our analysis becomes a kind of order parameter for
deconfinement. This is consistent with the lattice results~\cite{Boyd:1996ex}: $\langle G_{\mu\nu}^2\rangle$
is almost temperature-independent in the confined phase and drops sharply near the critical temperature
(strictly speaking, to negative values signaling instability of original
theory). In our 5D dual description, this effect is simulated as the Hawking--Page first
order phase transition from a gravitational background where the thermal AdS$_5$ space is
distorted by the dilaton to the AdS$_5$ space distorted by a black hole.

In order to make our analysis self-contained, in Section~2 we remind the reader the main steps of Herzog's
holographic calculation of deconfinement temperature in the HW model. Our derivation of $T_c$
is given in Section~3. Section~4 contains some discussions and phenomenological fits.
We conclude in Section~5.

\section{Deconfinement temperature in the HW model}

By assumption, the pure gravitational part of $5D$ action of holographic dual theory
for a 4D $SU(N)$ gauge theory has the form (the Euclidean signature is used)
\be
\label{1}
S = -\frac1{2\kappa^2}\int d^5x\sqrt{g} \left(R+\frac{12}{L^2} \right).
\ee
Here $g=\det g_{MN}$, $\kappa$ is the coefficient proportional to the $5D$ Newton constant,
$R$ is the Ricci scalar and $L$ represents the radius of AdS$_5$ space defined below.
The gravitational coupling scales as $\kappa \sim 1/N$.

The deconfinement in holographic models of QCD occurs as
the Hawking--Page phase transition between the following two gravitational backgrounds.
The first is the thermal AdS$_5$ space with a line element
\be
\label{2}
ds^2=\frac{L^2}{z^2}\left(d\tau^2+d\vec{x}^2+dz^2\right),
\ee
and the Euclidean time $\tau$ restrained to a finite interval $0\leq \tau\leq\beta$.
Here $z$ represents the holographic coordinate with physical meaning of inverse
energy scale.
The second background is the AdS$_5$ black hole\footnote{There are two different black holes
in the AdS space --- a "small" one and a "big" one~\cite{zwiebach}. A contribution of small black hole
to thermodynamics can be neglected~\cite{hawking}. One can show that the small one disappears at all
in the Poincar\'{e} patch~\eqref{2} of global AdS$_5$ space and one automatically deals with the
big black hole in~\eqref{3}.}
that describes the deconfined phase,
\be
\label{3}
ds^2=\frac{L^2}{z^2}\left(f(z)d\tau^2+d\vec{x}^2+\frac{dz^2}{f(z)}\right),
\ee
where $f(z)=1-(z/z_h)^4$ and $z_h$ denotes the horizon of the black hole.
The corresponding Hawking temperature is related to the horizon as $T=1/(\pi z_h)$.

The both solutions lead to the curvature $R = -20/L^2$.
Dividing out by the volume of $\vec{x}$ space one gets the free energy densities,
\be
\label{4}
V_1(\epsilon)=\frac{4L^3}{\kappa^2} \int_0^\beta d\tau\int_\epsilon^{z_m}\frac{dz}{z^5},
\ee
\be
\label{5}
V_2(\epsilon)= \frac{4L^3}{\kappa^2}  \int_0^{\pi z_h}d\tau\int_\epsilon^{\min(z_m,z_h)}\frac{dz}{z^5},
\ee
where $z_m$ represents the infrared cut-off of the HW model and an ultraviolet cut-off $z=\epsilon$
is introduced to regulate the arising infinity.
The two geometries are compared at $z=\epsilon$ where the periodicity in the
time direction is locally the same, {\it i.e.} $\beta=\pi z_h\sqrt{f(\epsilon)}$.
The order parameter for the phase transition is
\begin{equation}
\label{6}
\Delta V = \lim_{\epsilon\rightarrow0}\Delta V(\epsilon) = \lim_{\epsilon\rightarrow0}\left(V_2(\epsilon)-V_1(\epsilon)\right).
\end{equation}
The thermal AdS space is stable when $\Delta V>0$, otherwise the
black hole is stable. The condition $\Delta V=0$ defines the critical temperature $T_c$ at which the
transition between the two phases happens. It is easy to see that $\Delta V>0$ if $z_m<z_h$.
But if $z_m>z_h$ one has
\begin{equation}
\label{6b}
\frac{\kappa^2}{L^3\pi z_h}\Delta V(\epsilon)=\left(\frac{1}{\epsilon^4}-\frac{1}{z_h^2}\right)-
\left(\frac{1}{\epsilon^4}-\frac{1}{z_m^2}-\frac{1}{2z_h^2}\right).
\end{equation}
Thus the phase transition takes place at $z_m^4=2z_h^4$ corresponding to a temperature~\cite{Herzog}
\begin{equation}
\label{7}
T_c=\frac{2^{1/4}}{\pi z_m}.
\end{equation}
The numerical values depend on a choice of infrared cut-off $z_m$ which is usually
taken from the hadron phenomenology and is of the order of $1/\Lambda_{\text{QCD}}$.

\section{Deconfinement temperature in a solvable gravity-dilaton system}

The simplest extension of action~\eqref{1} to a gravity-dilaton system consists in
adding a dilaton kinetic term,
\be
\label{8}
S_d = -\frac1{2\kappa^2}\int d^5x\sqrt{g} \left(R+\frac{12}{L^2} - \frac12\partial_M\phi\partial^M\phi \right).
\ee
We assume that this action is dual to pure gluodynamics.
The backreaction of dilaton to the AdS$_5$ metric will describe holographically
the dynamical violation of scale invariance in gluodynamics leading to emergence of
positive gluon condensate.

Let us analyze the Hawking--Page phase transition between the gravitational background following from~\eqref{8}
and AdS$_5$ black hole. First of all we need the solution of Einstein equations for metric and dilaton profile.
The corresponding solution was found in Refs.~\cite{dilaton,dilaton2},
after continuation to Euclidean signature it takes the form
\be
\label{9}
ds^2=\frac{L^2}{z^2}\left(\sqrt{1-c^2z^8}\,\eta_{\mu\nu}dx^\mu dx^\nu+dz^2\right),
\ee
\be
\label{10}
\phi=\sqrt{\frac32}\log\frac{1+cz^4}{1-cz^4}+\phi_0,
\ee
where $c$ and $\phi_0$ are some constants. The behavior near the boundary $z\rightarrow0$,
$\phi=\sqrt{6}cz^4+\phi_0$, shows that according to AdS/CFT prescriptions~\cite{kleb} the constant
$c$ is proportional to the gluon condensate\footnote{We recall that the 5D mass of a scalar field $\phi$
dual to some 4D gauge theory operator $\mathcal{O}$ having the canonical dimension $\Delta$ is
$m_5^2 L^2 = \Delta(\Delta-4)$. The asymptotics of $\phi$ at $z\rightarrow0$ becomes
$\phi=c_1z^{4-\Delta}+c_2z^\Delta$, where up to renormalization constants $c_1$ corresponds to
the source of $\mathcal{O}$ and $c_2$ becomes proportional to v.e.v. $\langle\mathcal{O}\rangle$~\cite{kleb}.}
$\langle G_{\mu\nu}^2\rangle$.
The proportionality factor can be found from matching the leading term in the OPE of the gluon
operator $G_{\mu\nu}^2$ to the corresponding holographic calculation, the result in the
large-$N$ limit is~\cite{dilaton2}
\be
\label{11}
\langle G_{\mu\nu}^2\rangle=\frac{8\sqrt{3}\,N}{\pi}\,c.
\ee

The singularity at
\be
\label{12}
z_0=1/c^{1/4},
\ee
provides a natural infrared cut-off for the holographic coordinate which emerges dynamically.

Now we apply Herzog's analysis outlined in the previous Section to
the Hawking-Page phase transition between the AdS black hole geometry~\eqref{3} and the
thermal AdS dilaton-gravity geometry~\eqref{9} with the dilaton profile~\eqref{10}.
The two geometries are compared at $z=\epsilon$ where the periodicity in the
time direction is locally the same, {\it i.e.} $\beta=\pi z_h\sqrt{f(\epsilon)}/(1-c^2\epsilon^8)^{1/4}$.
In the limit $\epsilon\rightarrow0$, the correction from $\epsilon^8$ will not contribute,
hence, the comparison will be at $\beta=\pi z_h\sqrt{f(\epsilon)}$ as before.

The calculation of Ricci scalar for the geometry~\eqref{9} yields
\be
\label{14}
R=-\frac{20}{L^2}+\frac{48c^2z^8}{L^2(1-c^2z^8)^2}.
\ee
It is convenient to divide the contributions to free energy $V_1$ into three parts,
\be
\label{13}
V_1 = V_1^{(c)} + \Delta V_R + \Delta V_\phi,
\ee
where $V_1^{(c)}$ stems from the first term in~\eqref{14},
$\Delta V_R$ goes from the second one in~\eqref{14}, and
$\Delta V_\phi$ arises from the dilaton~\eqref{10}.
Using $\sqrt{g}=L^5(1-c^2z^8)/z^5$ from the geometry~\eqref{9}, we get
\be
\label{15}
\Delta V_R(\epsilon)=-\frac{24L^3c^2}{\kappa^2} \int_0^\beta d\tau\int_\epsilon^{z_0}\frac{z^3dz}{1-c^2z^8},
\ee
On the other hand, the solution~\eqref{9},~\eqref{10} gives
$\partial_M\phi\partial^M\phi=g^{zz}(\partial_z\phi)^2=96c^2z^4/\left(L^2(1-c^2z^8)^2\right)$,
that leads to
\be
\label{16}
\Delta V_\phi(\epsilon)=\frac{24L^3c^2}{\kappa^2} \int_0^\beta d\tau\int_\epsilon^{z_0}\frac{z^3dz}{1-c^2z^8}.
\ee
Comparing~\eqref{15} and~\eqref{16} we see that the last two contributions in~\eqref{13} cancel each other.
This important cancellation removes otherwise emerging logarithmic divergence
at
$\epsilon=0$.

The comparison of two geometries is thus reduced to comparison of free energies $V_2$ and $V_1 = V_1^{(c)}$
which is identical to the case of previous Section with the obvious replacement $z_m\rightarrow z_0=1/c^{1/4}$ in the
relation~\eqref{7},
\begin{equation}
\label{17}
T_c=\frac{(2c)^{1/4}}{\pi}.
\end{equation}
Inserting the renormalization factor~\eqref{11} we obtain our final result,
\be
\label{18}
T_c^4=\frac{\sqrt{3}\,\langle\frac{\alpha_s}{\pi}G_{\mu\nu}^2\rangle}{12\pi^2N\alpha_s},
\ee
where instead of scale-dependent gluon condensate $\langle G_{\mu\nu}^2\rangle$ we expressed $T_c$ via
an approximately renorminvariant quantity $\langle\frac{\alpha_s}{\pi}G_{\mu\nu}^2\rangle$
(the one-loop approximation to renorminvariant $\langle \beta G_{\mu\nu}^2\rangle$,
where $\beta=\beta(\alpha_s)$ denotes the QCD $\beta$-function)
which is usually extracted in QCD sum rules and lattice simulations. This
entails the appearance of additional parameter $\alpha_s$ which is
not known {\it a priori} since the energy scale is not fixed.

\section{Discussions and phenomenological fits}

An important technical point in our derivation was the cancellation
of the last two contributions in~\eqref{13}.
One can show that if in our analysis the gravity-dilaton system is
replaced by AdS$_5$ space with static dilaton background of standard SW holographic model~\cite{sw},
{\it i.e.} by action proportional to $\int d^5x\sqrt{g}\, e^{-cz^2} \left(R+12/L^2 \right)$,
then the given cancellation does not happen.
In the original Herzog's calculation of $T_c$ in the SW model~\cite{Herzog}, this problem was
avoided by ad hoc considering black hole in the same background $e^{-cz^2}$ although such a
gravitational solution is not known. This is another one troublesome point of analysis~\cite{Herzog}.

A qualitative correctness of relation~\eqref{18} can be motivated by a dimensional
analysis: If the gluon condensate $\langle \beta G_{\mu\nu}^2\rangle$
provides the only dimensional and renorminvariant scale in a theory then any dimensional and renorminvariant
quantity in this theory can be expressed as an appropriate power of
$\langle \beta G_{\mu\nu}^2\rangle$. Thus we should have
$T_c^4\sim\langle \beta G_{\mu\nu}^2\rangle$ by dimensionality.
We believe that such a natural relation should appear in any "natural" model
describing the deconfinement transition in a non-abelian gauge theory.
In this sense, the standard derivation of $T_c$ within the HW and SW
holographic models (and in their numerous successors) does not look "natural".
In the HW model, the infrared cut-off is not directly related to the gluon condensate
since non-perturbative corrections to the leading logarithm in two-point correlators
decrease exponentially. The required power-like corrections appear in the SW model and
this allows to relate the dimensional parameter of SW-like models to the gluon condensate.
But because of aforementioned artificial trick needed to cancel the ultraviolet divergence
in the difference of free energies, the resulting relation between $T_c$ and
gluon condensate takes a form of transcendental integral equation~\cite{Herzog}
which can be solved only numerically.

In order to make a phenomenological prediction from the relation~\eqref{18}
we will use the following procedure. The critical temperature
$T_c$ in the l.h.s. refers implicitly to pure gluodynamics, {\it
i.e.} it is $T_c^{\text{gl}}$ in the notation of Section~1. Let us
take the values of $T_c^{\text{gl}}$ and gluon condensate
$\langle\frac{\alpha_s}{\pi}G_{\mu\nu}^2\rangle$ in pure $SU(3)$ gluodynamics
from lattice simulations and fix thereby $\alpha_s$ for $N=3$.
To get an estimate for $T_c$ in the real word with physical quarks we will
substitute the value of $\langle\frac{\alpha_s}{\pi}G_{\mu\nu}^2\rangle$
from QCD sum rules. This can be partly justified in the large-$N$ limit
we are dealing with in holography: The quark effects are of the order of $O(N)$
while the gluon ones are of the order of $O(N^2)$, hence, the inclusion of
quarks in the fundamental representation should give $O(1/N)$ corrections
to~\eqref{18} which are beyond the validity of holographic approach.

We note in passing that on general grounds one expects a
decreasing of gluon condensate in presence of
quarks in comparison with pure Yang-Mills theory~\cite{svz}.
A similar decreasing was observed in lattice calculation for
$T_c$. As far as we know, earlier these two effects have not been related
in a manifest way. The obtained relation~\eqref{18} expresses and explains
the given proportionality.

The most quoted relevant lattice results for $SU(3)$ Yang-Mills
theory we found are: $T_c^{\text{gl}}=264$~MeV~\cite{Boyd:1996bx} and
$\langle\frac{\alpha_s}{\pi}G_{\mu\nu}^2\rangle=0.1$~GeV$^4$~\cite{camp}.
With these inputs we obtain from~\eqref{18} the value $\alpha_s=0.1$. This value of
QCD coupling roughly corresponds to a scale of $Z$-boson
mass~\cite{pdg}, where the perturbation theory in QCD becomes
robust. This might be an interesting prediction on its own.

A widely accepted phenomenological estimation of the gluon condensate in QCD
sum rules yields a smaller value
$\langle\frac{\alpha_s}{\pi}G_{\mu\nu}^2\rangle=0.012$~GeV$^4$~\cite{svz}.
Substituting this estimation to the relation~\eqref{18}, we obtain $T_c=156$~MeV
in remarkable agreement with the lattice and experimental values
mentioned in Section~1.

Finally we would mention a caveat about our analysis: The supergravity approximation 
is not trustful near the singularity\footnote{We would like to thank a referee for this remark.}~\eqref{12}. 
This does not seem to be a serious issue for the last two terms in the free energy~\eqref{13} which cancel each other
but in the first term the integral should extend to some uncertain IR cut-off $z_0=1/c^{1/4}-\delta$,
where $\delta>0$. This will lead to some positive contribution to the r.h.s. of~\eqref{18}.
The size of this contribution is not under our control. On the other hand, one might speculate
that this contribution is effectively absorbed by the value of $\alpha_s$ which we extracted ---
its original value in a corrected relation~\eqref{18} is larger than the estimated $\alpha_s=0.1$ but we rescale it
so as to include the contribution from the region where the supergravity approximation breaks down.
In other words, the relation~\eqref{18} without correction implies then that the uncertainty in question 
is effectively "hidden" in the employed phenomenological value of $\alpha_s$.

\section{Conclusions}

We proposed a new holographic calculation of deconfinement temperature at
vanishing chemical potential. Our calculation is free from some internal
inconsistencies inherent in many calculations of this sort starting from
Herzog's analysis~\cite{Herzog}. In our scheme, the phase transition from the hadron
phase to quark-gluon plasma is associated with a sharp
disappearance of positive definite gluon condensate $\langle G_{\mu\nu}^2\rangle$
in accord with lattice simulations. Motivated by the AdS/CFT correspondence,
as a holographic image of pure gluodynamics
we considered an exactly solvable gravity-dilaton system in which
gravity in AdS$_5$ space (by assumption, dual to a strongly coupled 4D conformal
gauge theory) is backreacted by a free massless dilaton field which is dual
via AdS/CFT prescriptions to source of gluon condensate.
The solution of corresponding Einstein equations is known to lead to a
singularity at some value of holographic coordinate that is
associated with dynamical cut-off describing holographically the violation of
scale invariance in QCD. From the analysis of Hawking-Page phase
transition between this gravity-dilaton system and black hole in AdS$_5$
(which is dual to deconfined phase in gauge theory) we got a
relation between the deconfinement temperature $T_c$ and gluon
condensate. After fixing a free parameter (gauge coupling $\alpha_s$)
from lattice data on $T_c$ and gluon condensate in $SU(3)$ pure
Yang-Mills theory, the derived relation~\eqref{18} yields $T_c=156$~MeV
for the accepted value of gluon condensate in QCD sum rules.
The obtained estimation of deconfinement temperature agrees perfectly
with the modern lattice data $156\pm9$~MeV~\cite{Borsanyi:2010bp} and
freeze-out temperature $T_{cf}=156.5\pm1.5$~MeV
measured by ALICE Collaboration at LHC~\cite{Andronic:2017pug}.

The proposed calculation of deconfinement temperature can be
applied in more contrived dynamical gravity-dilaton holographic
models (which include a dilaton potential and/or other scalar fields)
but most likely at the cost of a loss of exact analytical relations.
Also the effects of chemical potentials (normal, chiral and isospin)
can be considered.

\end{document}